\documentclass{jpp}
\usepackage{graphicx}
\usepackage{epstopdf, epsfig}
\usepackage{bm}
\usepackage[T1]{fontenc}

\shorttitle{Nonrelativistic pair shock}
\shortauthor{M. E. Dieckmann and A. Bret}

\title{Simulation study of the formation of a non-relativistic pair shock}

\author{M. E. Dieckmann\aff{1}
  \corresp{\email{mark.e.dieckmann@liu.se}},
 \and A. Bret\aff{2,3}}

\affiliation{\aff{1} Department of Science and Technology (ITN), Linkoping University, Campus Norrkoping, 60174 Norrkoping, Sweden
\aff{2}University of Castilla La Mancha, ETSI Ind, E-13071 Ciudad Real, Spain
\aff{3}Instituto de Investigaciones Energ\'eticas y Aplicaciones Industriales, Campus Universitario de Ciudad Real, 13071 Ciudad Real, Spain}

\begin{document}

\maketitle

\begin{abstract}
We examine with a particle-in-cell (PIC) simulation the collision of two equally dense clouds of cold pair plasma. The clouds interpenetrate until instabilities set in, which heat up the plasma and trigger the formation of a pair of shocks. The fastest-growing waves at the collision speed c/5 and low temperature are the electrostatic two-stream mode and the quasi-electrostatic oblique mode. Both waves grow and saturate via the formation of phase space vortices. The strong electric fields of these nonlinear plasma structures provide an efficient means of heating up and compressing the inflowing upstream leptons. The interaction of the hot leptons, which leak back into the upstream region, with the inflowing cool upstream leptons continuously drives electrostatic waves that mediate the shock. These waves heat up the inflowing upstream leptons primarily along the shock normal, which results in an anisotropic velocity distribution in the post-shock region. This distribution gives rise to the Weibel instability. Our simulation shows that even if the shock is mediated by quasi-electrostatic waves, strong magnetowaves will still develop in its downstream region.
\end{abstract}

\section{Introduction}

Compact objects like neutron stars or black holes that accrete material can emit relativistic jets. These jets are composed of electrons, positrons and ions. The emission of relativistic jets by microquasars \citep{Fabian79,Margon84}, which are stellar-size black holes that gather material from a companion star, and by some of the supermassive black holes in the centers of galaxies has been observed directly \citep{Bridle84}. The fireball model attributes gamma-ray bursts (GRBs) to ultrarelativistic jets, which are emitted during strong supernovae. A direct observation of the ultrarelativistic jets that trigger the GRBs and occur at cosmological distances is not possible. Their existence can thus not be established unambiguously \citep{Woosley06}. However, observations of a mildly relativistic plasma outflow during the supernova 1998bw by \citet{Kulkarni98} lend some support to the fireball model. 

The efficiency, with which the accreting object can accelerate the jet plasma, is not constant in time. A variable plasma acceleration efficiency results in a spatially varying velocity profile of the jet plasma. Internal shocks can form at locations with a large velocity change and these shocks can constitute strong sources of electromagnetic radiation \citep{Rees78}. The prompt emissions of gamma-ray bursts, which are associated with internal shocks in ultrarelativistic jets, are visible across cosmological distances and internal shocks should thus be sources of intense electromagnetic radiation. 

The relativistic factors of the internal shocks in GRB jets are probably of the order of a few. A wide range of theoretical and numerical studies have addressed the collision of lepton clouds at relativistic speeds and the instabilities that sustain the shock and thermalize the plasma that crosses it \citep{Kazimura98,Medvedev99,Brainerd00,Sakai00,Silva03,Haruki03,Jaroschek04,Medvedev05,Milos06,Chang08,Stockem08,Bret08,Bret2013,Sironi14,Marcowith16}. Such shocks thermalize plasma via the magnetic fields that are driven by the filamentation instability of counter-streaming beams of charged particles, which is also known as the beam-Weibel instability. 

This magnetic instability outgrows the competing electrostatic instabilities if two equally dense lepton clouds collide or interpenetrate at a relativistic speed. Electrostatic waves and instabilities can, however, not always be neglected and they exist even in lepton plasmas where positrons and electrons are equally dense. An external electric field accelerates electrons and positrons into opposite directions, which creates a current. Even if the initial electric field perturbation is removed, the current leads to a self-generated electric field in the plasma. The electric field eventually becomes strong enough to reverse the flow directions of electrons and positrons. The motion overshoots though and electrons and positrons oscillate around the equilibrium position. This oscillation is sustained by electrostatic fields, which can couple resonantly to other beams. 

The jets of microquasars contain a significant fraction of positrons \citep{Trigo13} alike the jets that trigger GRBs. The high variability of a microquasar jet, which is emitted by a stellar-size black hole that accretes material from a companion star, suggests that internal shocks are present in such jets \citep{Kaiser00,Miller05}. Microquasar jets expand at a lower speed than GRB jets. Typical velocity changes are thus likely to be smaller and the internal shocks of microquasars may not always be relativistic, in which case electrostatic processes may become more important. Nonrelativistic pair shocks have so far not received much attention and the structure of their transition layers remains unknown.

We explore with a particle-in-cell (PIC) simulation the initial evolution phase of a leptonic shock that forms when two equally dense pair clouds collide at a speed that is one fifth of the speed of light. The electrons and positrons of each cloud have an equal number density and mean speed and the net charge and current of each cloud vanishes. The thermal spread of the particle velocities is small compared to the cloud collision speed and the instabilities can initially be described in the cold plasma limit. Both, the pair temperature and the collision speed are probably too low to be realistic for the plasma distribution close to an internal shock of the jet of a microquasar. We chose these initial conditions because electrostatic instabilities tend to be more important than magnetic ones for nonrelativistic collisions of cold plasma. We can thus study a shock, for which electrostatic effects are maximized in its transition layer and that thus constitutes a lower bound for pair shocks with respect to the degree of magnetization. 

Our simulation shows that the transition layer of the pair shock is indeed mediated by nonlinear and predominantly electrostatic waves. However, the incomplete thermalization of the inflowing upstream plasma by the shock results in a downstream plasma with a thermally anisotropic distribution. This anisotropy is strong enough to trigger the growth of the Weibel instability in its original form \citep{Weibel59}. Even a practically electrostatic pair shock thus generates a magnetic field in its downstream region. The thermal anisotropy and, hence, the growth rate of the Weibel instability can be increased further by the presence of ions and by the ambipolar electric field they drive at the shock front \citep{Stockem14}.

Our paper is structured as follows. The shock formation mechanism, the PIC simulation method, our initial conditions and the expected spectrum of growing waves are discussed in Section 2. Section 3 presents the simulation results and section 4 is the summary.

\section{Shock formation, the simulation code and the initial conditions}

\subsection{The formation mechanism of a collisionless leptonic shock}
 
We examine the formation of shocks out of the collision of two charge- and current-neutral clouds of electrons and positrons. The plasma we consider is initially unmagnetized, no ions are present and all lepton species have the same temperature. The absent binary collisions imply that both lepton clouds will move through each other until plasma instabilities start to grow. Only three wave modes can develop for our initial conditions. The two-stream modes are purely electrostatic and their wave vector is aligned with the collision direction. The quasi-electrostatic oblique modes have a wave vector that is oriented obliquely to the collision direction and they belong to the same wave branch as the two-stream modes. The third mode is the filamentation mode, which is also known as the beam-Weibel mode \citep{Califano}. The wave vectors of these waves form the angle $\pi/2$ with the collision direction. These three modes grow simultaneously during the shock formation stage. Their growth is eventually halted by nonlinear processes, which heat up the plasma in the overlap layer and bring it closer to a thermal equilibrium. 

A leptonic shock can be created in a PIC simulation by the collision of one lepton cloud with a reflecting wall. The reflected leptons move against the inflowing leptons that have not yet reached the wall and an overlap layer develops. The instabilities in this overlap layer let waves grow that heat up the plasma when they saturate. The expansion of the heated plasma is limited on one side by the wall and a shock forms on the other side. The shock evolution is resolved correctly once a downstream region has formed that is thick enough to decouple the shock from the wall. The formation phase of the shock may, however, not be resolved correctly by this computationally efficient method. The mechanism that triggers the filamentation or beam-Weibel instability is that particles with oppositely directed current vectors repel each other and particles with parallel current vectors attract each other. The instability saturates by forming current channels that collect particles with the same direction of the current vector. Current channels that contain particles with oppositely directed current vectors are separated by magnetic fields. A reflection of a particle by the wall changes its velocity component along the wall's normal direction and, thus, the direction of its current vector. Spatially separated current channels can, however, not form at the wall because the particle is not spatially displaced by the reflection. The suppression of the filamentation instability at the reflecting wall will affect the spectrum of the unstable waves.

This spectrum is resolved correctly if we let two separate lepton clouds collide. If both clouds differ only in their mean speed, then we have to resolve in the simulation two identical shocks that enclose the expanding downstream region. It is computationally expensive and unnecessary to track both shocks for a long time. Here we let a long and a short lepton cloud collide. We increase the time interval during which we can observe the shock between the downstream region and the long lepton cloud. The second shock moves into the opposite direction and it eventually reaches the simulation boundary. By that time, it does no longer affect the evolution of the other shock. 

\subsection{The particle-in-cell (PIC) simulation method}

We model the collision of the lepton clouds with a particle-in-cell (PIC) simulation. The PIC simulation code is based on the kinetic plasma model, which approximates each plasma species $i$ by a phase space density distribution $f_i (\bm{x},\bm{v},t)$. The position vector $\bm{x}$ and the velocity vector $\bm{v}$ are treated as independent coordinates, which allows for arbitrary velocity distributions at any given position. The number density of this species is the zero'th moment of the distribution $n_i (\bm{x},t)=\int f_i (\bm{x},\bm{v},t) \, d\bm{v}$ and the mean speed $\bar{\bm{v}}_i (\bm{x},t) = \int \bm{v} f_i (\bm{x},\bm{v},t) \, d\bm{v}$ corresponds to its first moment. The number density and the mean speed yield the charge density $\rho_i (\bm{x},t) = q_i n_i (\bm{x},t)$ and the current density $\bm{J}_i (\bm{x},t) = q_i \bar{\bm{v}}_i (\bm{x},t) n_i (\bm{x},t)$ of the species $i$. The total charge density $\rho (\bm{x},t) = \sum_i \rho_i (\bm{x},t)$ and current density $\bm{J} (\bm{x},t) = \sum_i \bm{J}_i (\bm{x},t)$ update the electromagnetic fields via a finite difference approximation of Ampere's and Faraday's laws on a numerical grid.
$$
\mu_0 \epsilon_0 \frac{\partial \bm{E}}{\partial t} = \nabla \times \bm{B} - \mu_0 \bm{J}, 
$$
$$
\frac{\partial \bm{B}}{\partial t} = -\nabla \times \bm{E}.
$$
The EPOCH code \citep{Arber15} we use fulfills $\nabla \cdot \bm{B}=0$ and $\nabla \cdot \bm{E} = \rho / \epsilon_0$ to round-off precision. 

An ensemble of computational particles (CPs) with the charge $q_i$ and mass $m_i$ approximates the phase space density distribution $f_i (\bm{x},\bm{v},t)$. The relativistic momentum $\bm{p}_j$ of the $j^{th}$ CP of species $i$ is updated via a discretized form of the relativistic Lorentz force equation $\frac{\partial }{\partial t} \bm{p}_j = q_i \left ( \bm{E}(\bm{x}_j) + \bm{v}_j \times \bm{B}(\bm{x}_j)\right )$ and its position is updated via $\frac{\partial}{\partial t} \bm{x}_j = \bm{v}_j$. The electric field and the magnetic field are interpolated from the numerical grid to the particle's position $\bm{x}_j$ to update its momentum. The current density on the grid, which is used to update the electromagnetic fields, is the sum over all particle currents after they have been interpolated from the particle positions to the grid nodes.

\subsection{The simulation setup}

Our two-dimensional simulation box has the length $L_x$ along $x$ and $L_y$ along $y$. The simulation box is subdivided into the two intervals $-0.65 \, L_x < x < 0$ and $0 \le x < 0.35 \, L_x$. The boundary at $0.35\, L_x$ is reflecting and that at $-0.65 \, L_x$ is open. The boundary conditions at $y=0$ and $y=L_y$ are periodic. We place electrons and positrons with equal densities $n_0$ and temperatures $T_0$ = 10 eV everywhere in the box at $t=0$. The electrons and positrons in the interval with $x\ge 0$ have a vanishing mean speed. The electrons and positrons in the interval $x<0$ have the mean speed $v_0 = 0.2c$ along $x$. No new particles are introduced while the simulation is running and the simulation is stopped well before the end of the inflowing lepton cloud encounters the shock or before the leptons that are reflected by the boundary at $x=0.35 \, L_x$ return to the shock. 

We normalize the position to the electron skin depth $\lambda_s = c / \omega_p$, where $\omega_p = {(n_0 e^2/\epsilon_0 m_e)}^{1/2}$ is the electron plasma frequency of one cloud. Velocities are normalized to $c$. Momenta are normalized to $c m_e$ and we define $p_0 = v_0 m_e$ as the mean momentum of a lepton of the plasma cloud in the half-space $x<0$. The box size $L_x \times L_y = 60 \times 2.4$ is resolved by $1.9 \times 10^4$ grid cells along $x$ and by 760 grid cells along $y$. Electrons and positrons are represented by 25 CPs per cell, respectively. The time is normalized to $\omega_p^{-1}$. The simulation time $t_{sim}=120$, which is subdivided into $57200$ equal time steps. We normalize the electric field to $\omega_p c m_e / e$ and the magnetic field to $\omega_p m_e / e$.

\subsection{The solution of the linear dispersion relation}

We have to verify that our box is large enough to resolve the competing unstable modes and we want to determine the wave mode, which grows fastest for the selected initial conditions. We solve for this purpose the linear dispersion relation in order to determine the spectrum of the growing waves. The solution is computed under the assumption that the overlap layer has an infinite size. This condition is approximately fulfilled if the colliding clouds can interpenetrate for some time before the instabilities grow.

The initial velocity spread for $T_0=10$ eV is about $v_{th0}=4.5 \times 10^{-3}c$ and both clouds drift toward each other at $v_0=0.2c$. Thermal effects can be neglected for the ratio $v_0/v_{th0} = 44$ and the lepton beams are cold. We solve the linear dispersion relation in the frame of reference in which the total momentum vanishes. The pair clouds move in this reference frame into opposite x-directions at the speed modulus $\beta_0'\equiv v_0'/c=1/10$. 

The non-relativistic dispersion equation for a perturbation of the form $\exp (i \bm{k}\cdot \bm{r} - i\omega t)$ and a wave vector $\bm{k}$ with an arbitrary orientation is \citep{Bret10}
\begin{equation}\label{eq:dispersion2}
\left(\omega^2\epsilon_{xx} -  k_y^2  c^2\right)\left(\omega^2\epsilon_{yy} - k_x^2  c^2\right)
- \left(\omega^2\epsilon_{yx} + k_xk_y c^2\right)^2 = 0 \,.
\end{equation}
where $\delta _{\alpha \beta }$ is the Kronecker symbol and
\begin{equation}\label{eq:epsi_general}
    \epsilon_{\alpha \beta }(\bm{k},\omega) = \delta _{\alpha \beta }\left(1- \frac{\omega_{p}^2}{\omega^2} \right)
+\frac{\omega_{p}^2}{\omega^2}\sum_j\int d^3p\,
\frac{p_{\alpha }p_{\beta } \bm{k}\cdot \left(\frac{\partial f_j^0}{\partial \bm{p}}\right)}{m\omega -\bm{k}\cdot \bm{p} } .
\end{equation}
The problem of finding the fastest growing mode has been solved \citep{Bret2005,Bret2013} for cold distributions of the form $f^0_j(\bm{p})=\delta(p_y)\delta(p_x-P_j)$. 

\begin{figure}
{\includegraphics[width=0.6\columnwidth]{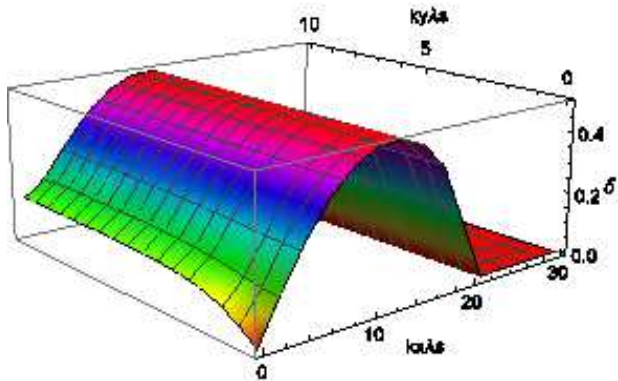}}
\caption{The solution of the linear dispersion relation for two beams, each of which consists of electrons and positrons with the same number density, mean speed and temperature $T_0 = $ 10 eV. The beams have an infinite extent and they counterstream along $x$ with the speed modulus $0.1c$. The growth rate $\delta$ is expressed in units of $\omega_p$.}\label{Disper}
\end{figure}

Figure \ref{Disper} shows the solution of the linear dispersion relation for our plasma parameters.
The growth rate peaks at the wave number $k_{x}\lambda_s \approx 12$ and its value does not depend on $k_y \lambda_s$ for the considered wave number interval. The fastest-growing modes are thus the two-stream/oblique modes. Their peak exponential growth rate is
\begin{equation}\label{eq:TS}
\frac{\delta_{TS}}{\omega_p} = \frac{\sqrt{2}}{2}.
\end{equation}
The filamentation modes are characterized by a flow aligned component $k_x = 0$. \citet{Califano} estimated their growth rate as
\begin{equation}\label{eq:Weibel}
\frac{\delta_W}{\omega_p} =2\beta_0'.
\end{equation}
Figure \ref{Disper} demonstrates that the growth rate of the filamentation modes with $k_x = 0$ is smaller than that of the two-stream/oblique modes, which confirms the aforementoned approximations since $\delta_W < \delta_{TS}$ for $\beta_0'=1/10$. 

We can estimate with the help of Fig. \ref{Disper} if and how our limited box size will affect the spectrum of growing waves. The simulation employs periodic boundary conditions along $y$ and the box length is $L_y$ in this direction. The smallest resolved wave number is thus $k_c = 2\pi / L_y$ or $k_c\lambda_s = 2.6$ and waves with $k_y < k_c$ can not grow. Figure \ref{Disper} shows that the growth rate of the filamentation modes decreases below $\delta_W$ for $k_y < k_c$ while that of the two-stream/oblique modes remains unchanged. The main effect of the limited box size along $L_y$ is thus to suppress the wave numbers where the growth of the filamentation instability is negligible. If our simulation shows that the plasma dynamics is governed by the two-stream/oblique modes, then we would obtain the same result also for larger $L_y$.

\section{Simulation results}

We discuss the simulation results at selected times and focus on the shock that forms at lower values of $x$. The first part addresses the wave modes that trigger the formation of shocks. The second part discusses the structure of the shock and the electric fields that mediate it and the final part examines the growth of magnetic fields.

\subsection{Instability and nonlinear saturation}

\begin{figure}
\centerline{\includegraphics[width=\columnwidth]{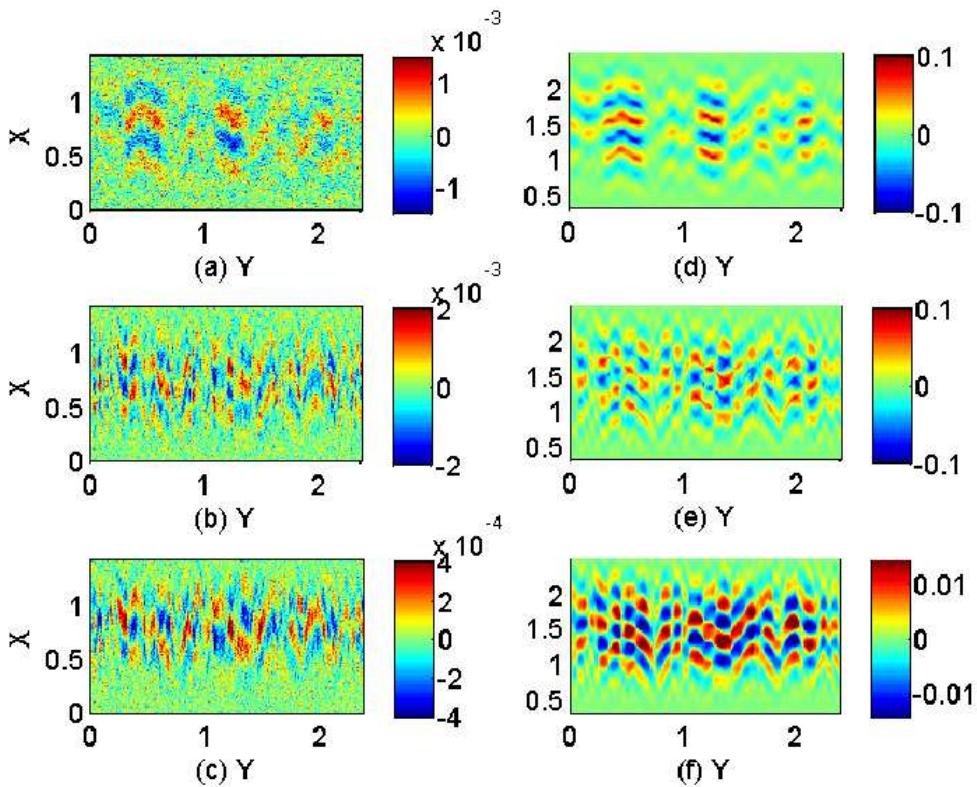}}
\caption{The in-plane electric field and the out-of-plane magnetic field close to the initial collision boundary at the time $t_1 = 7.6$ (left column) and at $t_2 = 14.7$ (right column): Panels (a, d) show $E_x$, panels (b, e) show $E_y$ and panels (c, f) show $B_z$.}\label{FieldsStep1836}
\end{figure}

The two clouds of initially unmagnetized collisionless lepton plasma will move through each other for some time before plasma instabilities set in. Figure \ref{FieldsStep1836}(a-c) displays the electric $E_x$ and $E_y$ components as well as the magnetic $B_z$ component at the time $t_1=7.6$. 
The lepton cloud, which was initially located in the half-space $x<0$, has moved by $v_0 t_1 = 1.5$ towards increasing values of x. Waves have grown in the cloud overlap layer, which spans the interval $0<x<1.5$ at this time. The distribution of $E_x$ reveals waves with a wavelength $\lambda \approx 0.4$. The $E_y$ and $B_z$ components are closely correlated and both oscillate rapidly along y. 

The in-plane electric field components and the out-of-plane magnetic field at the time $t_2=14.7$ are displayed in Fig. \ref{FieldsStep1836}(d-f). Figures \ref{FieldsStep1836}(a) and (d) show the same distribution of $E_x$ except for the larger amplitude. Their spatial confinement demonstrates that these waves do not propagate along $y$. The wave structures belong to electrostatic two-stream modes. The patterns in $E_y$ resemble those in $B_z$ and their amplitude ratio is comparable to that at the earlier time. The spatial correlation of the field structures in the distribution of $E_y$ and $B_z$ suggests that they belong to the same waves. 

We can extract some properties of the waves from a comparison of the amplitude of $E_y$ and $B_z$ at the times $t_1$ or $t_2$. The ratio of the field energy densities $\epsilon_0 (E_x^2 + E_y^2) / 2$ and $B_z^2/2\mu_0$ is in the given normalization $(E_x^2 + E_y^2)/ B_z^2 \approx 100$. The particles of both clouds move at a speed $\approx v_0/2$ relative to the waves, which are slow-moving in the reference frame of the overlap layer. The electric force imposed on a charged particle, which moves with $v_0/2$ = 0.1, is 50 times larger than the magnetic force. We conclude that the wave's magnetic field neither has a significant energy density nor does it affect the lepton dynamics. The waves are thus quasi-electrostatic and their wavelength along the collision direction is $\approx 0.4$. The amplitude of the waves has increased by a factor $\approx 50$ during the time interval $t_2-t_1 = 7.1$. If we assume that the waves grow exponentially, then their growth rate is $\delta \approx 0.5$ in units of $\omega_p$, which matches that in Fig. \ref{Disper}. 

\begin{figure}
{\includegraphics[width=0.5\columnwidth]{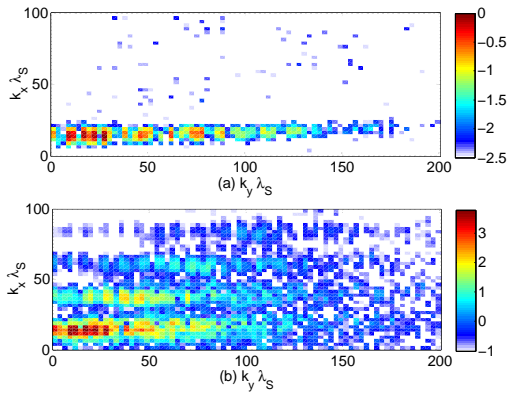}}
\caption{The spatial power spectra of the in-plane electric field $E_p = E_x(x,y) + iE_y(x,y)$ at the time $t_1 = 7.6$ (a) and $t_2 = 14.7$ (b). The color scale is 10-logarithmic and both spectra are normalized to the peak value in (a).}\label{FourierStep1836}
\end{figure}

The wave modes that yield the observed electric field can be identified with its spatial power spectrum. We Fourier-transform the in-plane electric field distribution $E_p(x,y) = E_x(x,y) + iE_y(x,y)$ over the spatial interval $0.2 < x < 2.7$ and over all $y$ and multiply it with its complex conjugate. Figures \ref{FourierStep1836}(a, b) show the power spectra at the times $t_1$ and $t_2$ in the quadrant $k_x > 0$ and $k_y > 0$.
The power spectrum at $t_1 = 7.6$ shows wave power at $k_x \lambda_s \approx 14$, which extends up to maximum perpendicular wave number $|k_y \lambda_s| \approx 200$. The wavenumber $k_x \lambda_s =14$ corresponds to a wavelength along $x$ of about $0.45$. 

The flow-aligned wave number $k_x \lambda_s \approx 14$ of the fastest-growing waves and the extension of wave power to large values of $k_y$ agree with the numerical solution of the linear dispersion relation in Fig. \ref{Disper}. The solution of the linear dispersion relation predicts a peak growth rate that does not depend on the value of $k_y$ for the considered wave numbers. The wave spectrum on Fig. \ref{FourierStep1836}(a) does however suggest that waves with a low value of $k_y$ grow faster. The growth rate is proportional to the amplitude the wave would reach after a given time if its growth would not be limited by nonlinear effects. The electric field amplitude, which is necessary to form phase space vortices, decreases with increasing values of $k = {|k_x^2 + k_y^2|}^{1/2}$ \citep{ONeil65} and the discrepancy between the spectral distribution in Fig. \ref{FourierStep1836}(a) and the solution of the linear dispersion relation in Fig. \ref{FourierStep1836}(c) can thus be explained with a stronger nonlinear damping that is imposed on the modes with a large value of $k_y$. 

The power spectrum in Fig. \ref{FourierStep1836}(b) is still concentrated on the two-stream / oblique mode branch. Its width along $k_y$ has diminished, which suggests that thermal damping is at work; the range of wave numbers $k_y$ that are unstable to the oblique mode instability is large in a cold plasma, while the wave growth is concentrated at low values of $k_y$ if the plasma is hot \citep{Silva2002}. A first and second harmonic along $k_x$ have emerged. The wave amplitudes have thus reached a non-linear regime \citep{Umeda03}. 

\begin{figure}
 \centerline{\includegraphics[width=0.5\columnwidth]{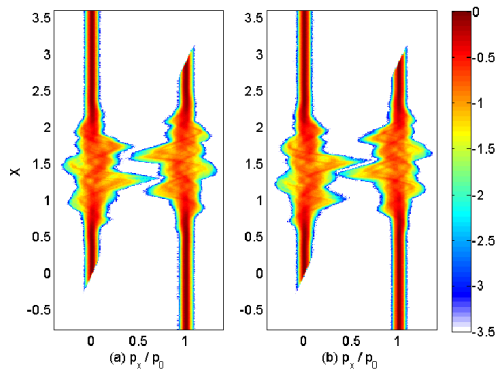}}
\caption{The phase space density distribution in the $x,p_x$-plane at the time $t_2=14.7$ of electrons (a) and positrons (b). The phase space density distribution is averaged over all other dimensions. The momentum is normalized to $p_0$. The color scale is 10-logarithmic.}\label{xpxStep36}
\end{figure}

Non-linear effects in the wave distribution should be tied to changes in the lepton distribution. Figure \ref{xpxStep36} shows the phase space density distributions $f(x,p_x)$ of the electrons and positrons. 
The overlap layer of both clouds spans the interval $-0.2 < x < 3$. The counterstreaming clouds have not yet merged along $p_x$. However, the substantial particle acceleration demonstrates that the instability is about to saturate. The density in the overlap layer is twice that of a single cloud and the density flucutations caused by the waves are of the order of $5\% - 10\%$ (not shown).

The filamentation instability starts to grow immediately but two-stream outgrows it.

\subsection{Shock formation}

The two-stream instability saturates by forming stable phase space vortices in the electron and positron distributions \citep{Berk67} and the same holds during the initial saturation stage of the oblique mode instability \citep{Dieckmann06}. Electron phase space vortices are characterized by strong bipolar pulses in the electric field distribution, which correspond to a localized positive excess charge. Positron phase space vortices correspond to a localized negative excess charge.

\begin{figure}
{\includegraphics[width=0.5\columnwidth]{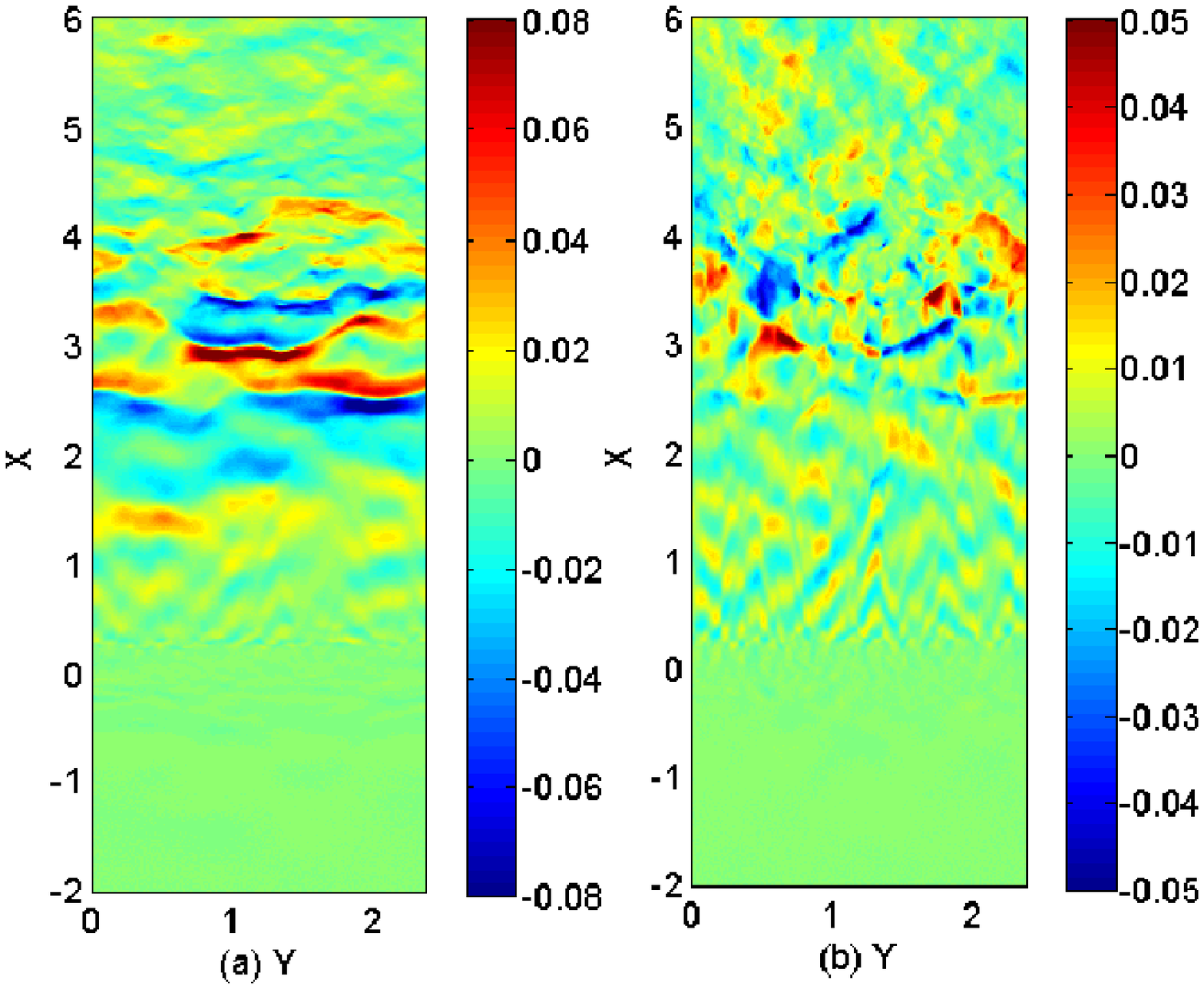}}
{\includegraphics[width=0.5\columnwidth]{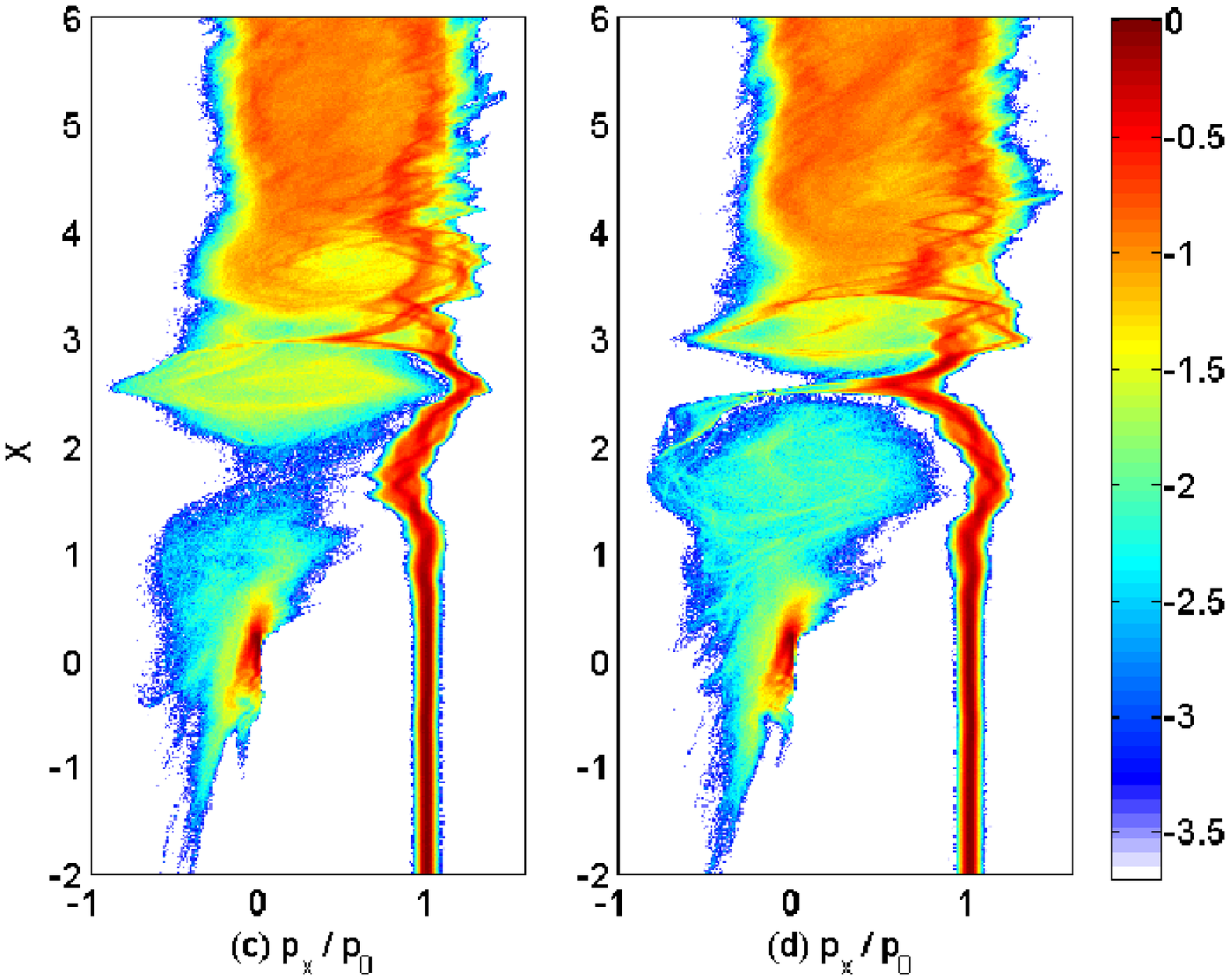}}
\caption{Panels (a) and (b) show the electric $E_x$ and $E_y$ components close to the initial collision boundary. Panels (c) and (d) show the phase space density distributions in the $x,p_x$-plane of electrons and positrons, respectively. The phase space density distributions are averaged over all other dimensions, they are normalized to the same value and displayed on a 10-logarithmic scale. The momentum is normalized to $p_0$. The simulation time is $t_3=58.8$.}\label{FieldsStep144}
\end{figure}

The in-plane electric field components at the time $t_3=58.8$ are displayed in Fig. \ref{FieldsStep144}(a,b).
The electric $E_x$ component shows such bipolar field structures. A large quasi-planar field pulse is located at $x\approx 2.5$ in the interval starting from $y\approx 1$ that goes through the periodic boundary at $y=2.4$ until $y\approx 0.5$. The polarity of $E_x$ indicates the presence of a positive excess charge in between both electric field bands. If this quasi-planar bipolar pulse is associated with an electron phase space vortex, then the latter should be detectable in the electron phase space density distribution even if it has been integrated over all values of $y$. 

Figures \ref{FieldsStep144}(c, d) show the corresponding electron and positron distributions. Figure \ref{FieldsStep144}(c) confirms the existence of a phase space vortex in the electron distribution at this location. The vortex in Fig. \ref{FieldsStep144}(c) spans the spatial interval $2<x<3$ and the momentum interval $-1 < p_x / p_0 < 1$. The mean momenta of the upstream electrons and positrons are modulated by the electrostatic potential of the vortex when they pass it, but they are not trapped by it. The upstream leptons continue to move to increasing values of $x$ until they are thermalized upon entering the downstream region $x>4$, which is characterized by a dense phase space density distribution between $0 < p_x/p_0 < 1$. This thermalization can only be accomplished by the field structures seen in the in-plane electric field between $x \approx 3$ and $x\approx 4$ in Fig. \ref{FieldsStep144}(a, b). 

The distribution of the positrons shows two smaller vortices that surround the large electron phase space vortex. The positron vertices are centered at $x\approx 1.8$ and $x\approx 3$. The zero-crossing of the electric $E_x$ component and, thus, the extremal point of the electrostatic potential at $x\approx 2.5$ in Fig. \ref{FieldsStep144}(a) corresponds to a stable equilibrium point for the trapped electrons. Hence it is an unstable equilibrium point for the positrons, explaining why the vortices of positrons and electrons are staggered along $x$. 

A small localized cloud of electrons and positrons is centred at $x \approx 0$ and $p_x \approx 0$. The cloud is an artifact from our initial conditions. The finite growth time of the electrostatic instabilities implies that the waves start to grow well behind the front of the plasma cloud that was initially located in the half-space $x>0$. This charge- and current neutral cloud is stable against electrostatic instabilities, because its extent along $x$ is not sufficiently large to allow it to interact with the inflowing upstream leptons via a two-stream instability.

\begin{figure}
{\includegraphics[width=0.5\columnwidth]{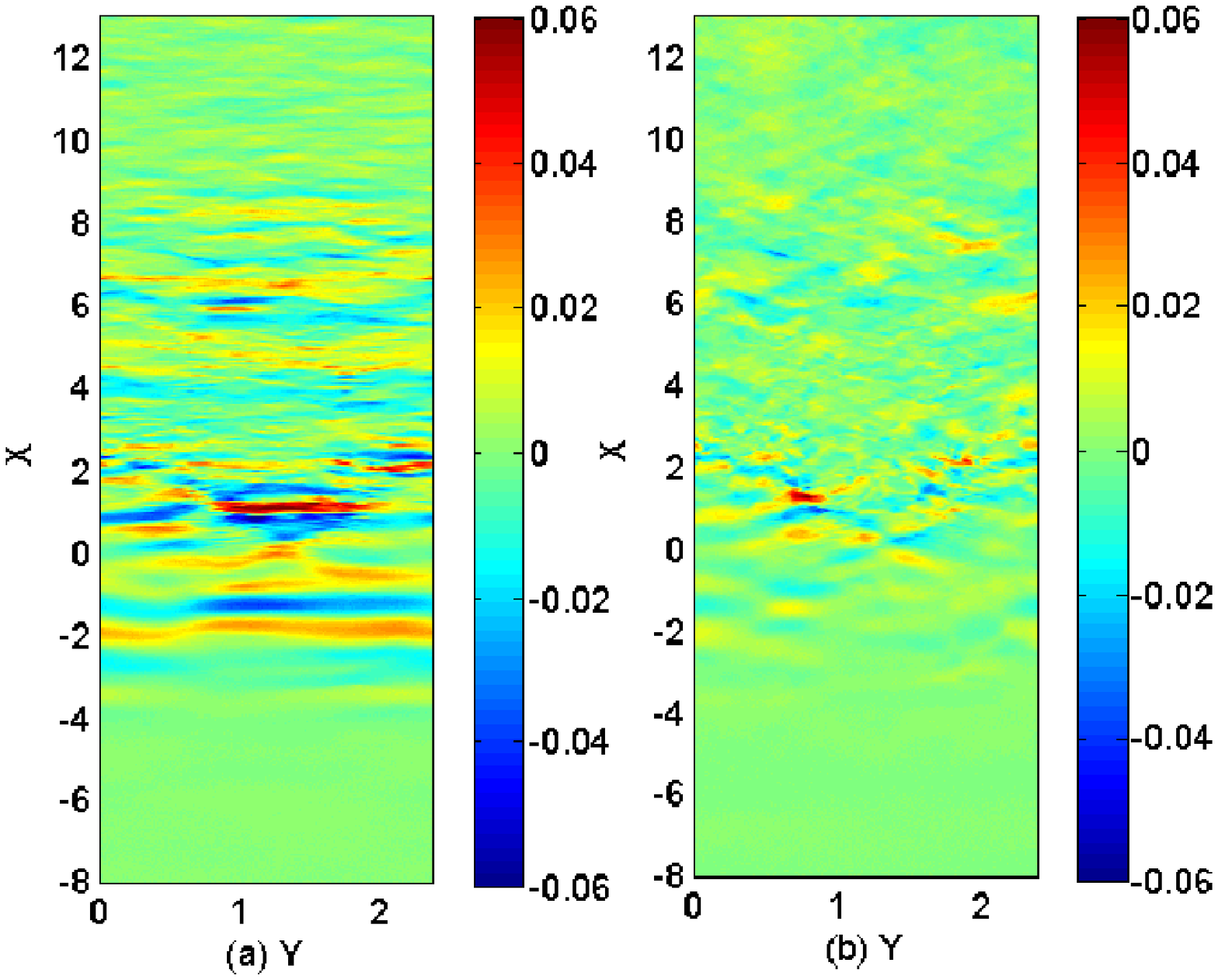}}
{\includegraphics[width=0.5\columnwidth]{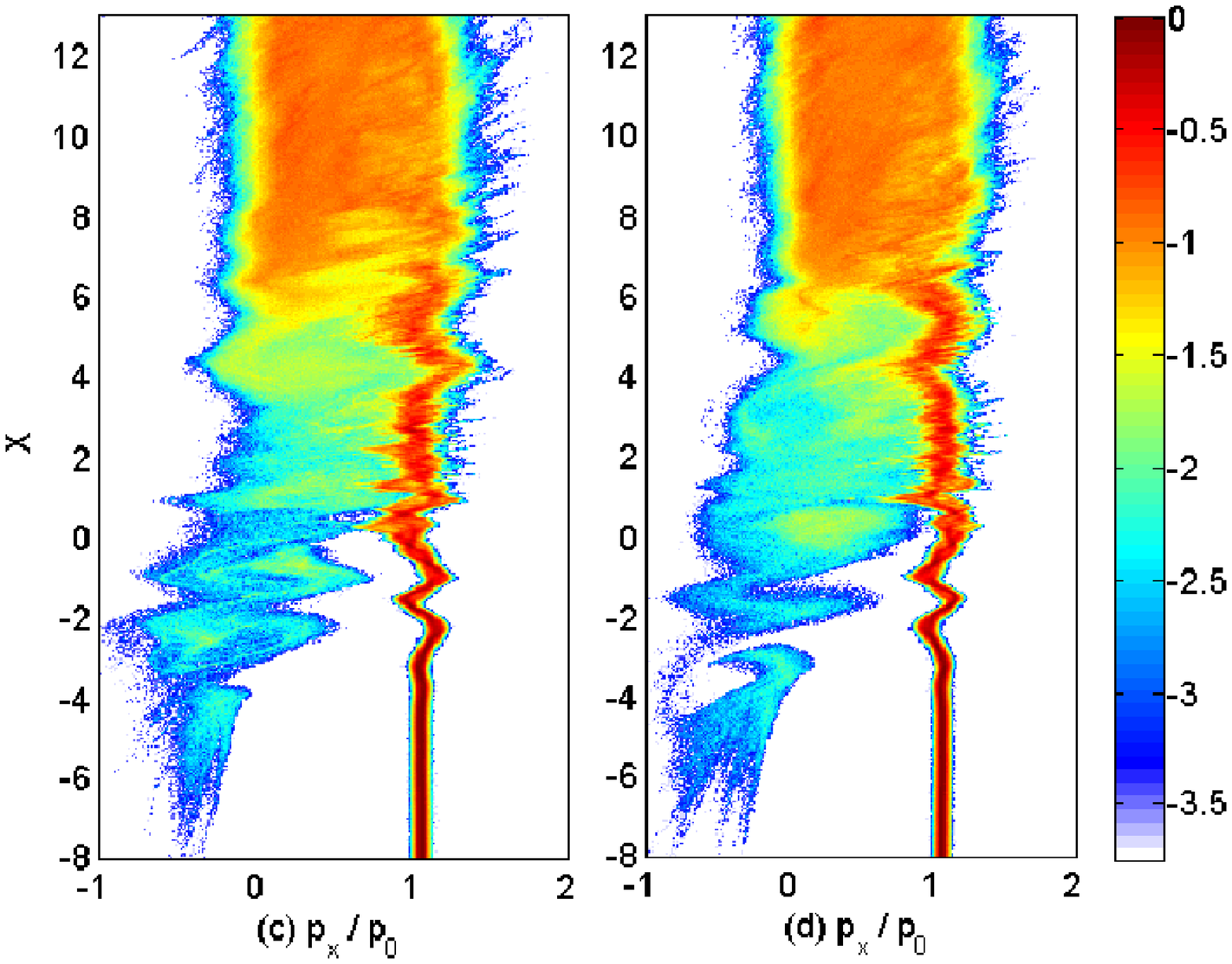}}
\caption{Panels (a) and (b) show the electric $E_x$ and $E_y$ components close to the initial collision boundary. Panels (c) and (d) show the phase space density distributions in the $x,p_x$-plane of electrons and positrons, respectively. The phase space density distributions are averaged over all other dimensions, they are normalized to the same value and displayed on a 10-logarithmic scale. The momentum is normalized to $p_0$. The simulation time is $t_4=120$.}\label{FieldStep297}
\end{figure}

 
Figure \ref{FieldStep297} shows the in-plane electric field distribution and the associated lepton phase space density distributions at the time $t_4=120$.
We observe strong quasi-planar electric field structures in the $E_x$-distribution in the interval $-3 < x <0$. Their amplitude is comparable to the one that gave rise to phase space vortices in the electron- and positron distributions at the earlier time. These electrostatic structures in $E_x$ have propagated well beyond the initial collision boundary $x=0$ reaching a position $x\approx -3$. We find relatively strong electric field oscillations in $E_x$ and $E_y$ between $0 < x < 8$. The transition layer of this shock thus spans at this time an interval with the width $\Delta x \approx 10$.

The strong planar waves in the interval $-3 < x < 0$ in Fig. \ref{FieldStep297} are correlated with phase space vortices in the hot lepton population at low speeds. The vortices of electrons and positrons are staggered along $x$. The electrons and positrons that gyrate in these vortices originate from the hot plasma component and they are well-separated along $p_x$ from the inflowing upstream leptons. The mean speed of these phase space vortices is less than $p_x = 0$, which implies that they move towards decreasing values of $x$. The mean speed of the vortices decreases with an increasing distance from the shock transition layer and they are thus accelerated away from the downstream plasma. The leptons, which gyrate in the vortices, reach a peak momentum $\approx -p_0$. 

The simultaneous presence in the interval $-6 < x < 5$ of the hot leptons that have leaked from the downstream region and the cooler drifting upstream leptons implies that the overall plasma distribution is non-thermal and thus unstable. The electric field of the phase space vortices seeds the instability and we observe momentum oscillations along $p_x$ in the cool inflowing electrons and positrons that increase with $x$ in the interval $-3 < x < -1$. The oppositely directed oscillations of electrons and positrons result in a strong current, which induces an electric field. The electric field oscillates in space and its oscillation amplitude decreases in unison with the net current in the direction of decreasing values of $x$. We can describe this oscillation in terms of a product between a sinusoidally oscillating electric field and an envelope function. 

A spatially varying envelope function gives rise to a ponderomotive force \citep{Kono80} that does not depend on the sign of the particle charge and accelerates electrons and positrons in the direction of decreasing values of the envelope function. This ponderomotive force is excerted by the modulated upstream plasma onto the hot leptons that form the phase space vortices and it accelerates them.

The leptons in the interval $1 < x < 6$ in Fig. \ref{FieldStep297}(c,d) are composed of a hot dilute component and the cool dense upstream leptons. Both populations gradually mix and they merge to a single one at $x \approx 6$. We observe electric fields in this interval in Fig. \ref{FieldStep297}. These fields show some piecewise planar structures, which correspond to phase space vortices with a limited extent along $y$. The two strongest localized structures at $x\approx 1$ are separated by a perpendicular $E_y$ field at $y \approx 0.8$. These localized structures are likely to be the result of an instability of initially planar phase space vortices or phase space tubes. Indeed, two-dimensional PIC simulations \citep{Oppenheim99} of phase space tubes in a stabilizing magnetic field show that the phase space tubes gradually break up along their axes. The collapse of a phase space vortex is an effective way to scatter the leptons in phase space, which results in a mixing of the downstream and upstream leptons in the $x,p_x$ plane. Strong small-scale electric fields are observed up to $x\approx 8$. The absence of phase space vortices with $x > 8$ demonstrates that the lepton distribution in this interval is no longer unstable to electrostatic instabilities.

\begin{figure}
 \centerline{\includegraphics[width=0.5\columnwidth]{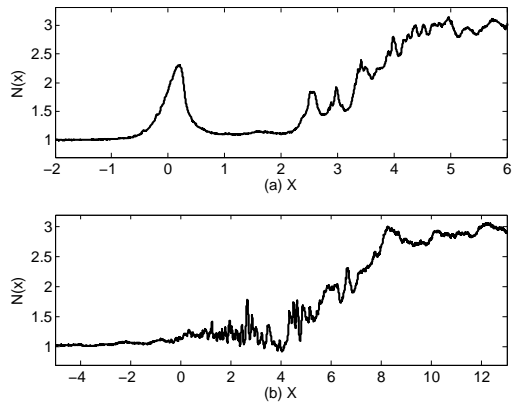}}
\caption{The total lepton density $N(x)$ in units of the initial total density $2n_0$ at the time $t_3 = 58.8$ (a) and $t_4 = 120$ (b).}\label{Density} 
\end{figure}

The phase space density distribution of the leptons was uniform in the interval $4 < x < 6$ at the time $t_3 = 144$ and in the interval $7 < x < 13$ at the time $t_4 = 297$. We conclude that these intervals correspond to a downstream region that is close to being in a thermal equilibrium, at least with respect to electrostatic waves and instabilities. The density distribution along $x$ sheds light onto how much the plasma is compressed by the shock crossing. Figure \ref{Density} compares the density distributions at the times $t_3$ and $t_4$.
The density converges at low $x$ to the initial density. A density peak is observed close to $x\approx 0$ in Fig. \ref{Density}(a), which corresponds to the dense lepton cloud at this position shown in Fig. \ref{FieldsStep144}(a). The density peak has disappeared in Fig. \ref{Density}(b). The electric fields that grew in this spatial interval in response to the instability between the inflowing upstream leptons and the leaked hot downstream leptons teared the cloud apart by accelerating the electrons and positrons into opposite directions. The density rises from about $N(x)=1$ to $N(x) \approx 3$ over a few electron skin depths. The plasma compression factor of about 3 is the one expected for a strong nonrelativistic shock \citep{Zeldovich67}.

\subsection{Secondary instabilities and magnetic field generation}

\begin{figure}
\centerline{\includegraphics[width=0.6\columnwidth]{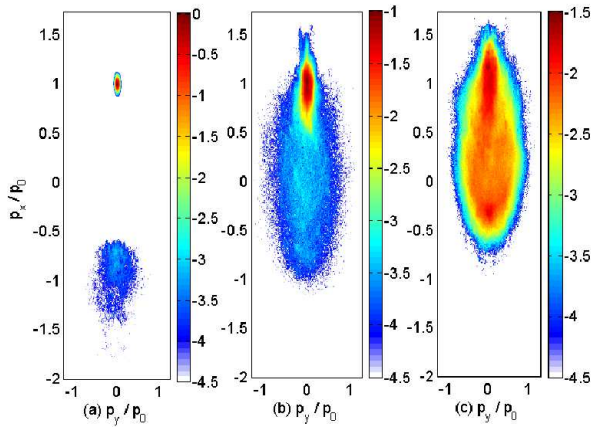}}
\caption{The total lepton distribution $f_t (v_x,v_y)$ averaged over $-4.5 < x < -4.2$ is shown in panel (a), that averaged over $1.9 < x < 2.2$ in panel (b) and that averaged over $8.3 < x < 8.6$ is shown in panel (c). The color scale is 10-logaritmic and normalized to the peak value in (a). }\label{vxvy}
\end{figure}

The electric fields associated with the phase space vortices heat up the leptons via Landau damping \citep{Landau46,ONeil65} and their collapse scatters them in phase space. The effects of this heating on the lepton distribution is visualized by Fig. \ref{vxvy}, which shows the phase space density as a function of $p_x$ and $p_y$ at three positions along $x$. The distribution has been integrated over $y$ and over an interval along $x$ of width 0.3.

The distribution in Fig. \ref{vxvy}(a) has been sampled far upstream of the shock. The upstream leptons constitute the cold dense beam that is located at $p_x \approx p_0$. The leaked leptons form a hot and dilute beam that moves at $p_x \approx -p_0$. The mean speed of the hot lepton beam exceeds that expected from a specular reflection, since the shock is moving to increasing values of $x$. Figure \ref{vxvy}(b) reveals that the inflowing upstream leptons have been heated up by the time they reach the position $x \approx 2$. They are distributed over a wider velocity range and their peak value of the phase space density has thus decreased. The temperature is of the order of 100 eV. These leptons are immersed in a hot dilute lepton component. Its thermal momentum spread is of the order of $p_0$ and the temperature is thus about one keV. The inflowing upstream leptons form a hot beam at $p_x \approx p_0$ in \ref{vxvy}(c) that is only about twice as dense as the leptons in the hot population. 

The waves observed close to $x=2$ in Fig. \ref{FieldStep297}(a,b) suggest that the velocity distribution in Fig. \ref{vxvy}(b) is still unstable to an electrostatic instability. It can not be the two-stream instability because that one requires two beams that are well-separated along $p_x$. This distribution can, however, still be unstable to the electron acoustic instability. Alike the well-known ion acoustic instability, which is driven by a drift between cold ions and hot electrons, the electron acoustic instability can develop if cold electrons drift relative to a hot electron species. Waves grow if the drift speed between the hot and the cold electron species exceeds several times the thermal spread of the cold electron species \citep{Gary87}. This condition is fullfilled in Fig. \ref{vxvy}(b). We note in this context that although the phase space density of the hot leptons is two orders of magnitude less than that of the inflowing upstream leptons their number density, which we obtain by integrating the phase space density along $p_x$, is of the same order. The interaction of counterstreaming lepton beams with a similar density results in rapidly growing instabilities. To the best of our knowledge the acoustic instability in pair plasma has not yet been explored. Here we can not unambigously show that it exists in pair plasma, because the electric field may also be the residual field of a phase space hole that formed previously.

The distribution in Fig. \ref{vxvy}(c) appears to be stable against electrostatic instabilities since we do not observe significant electric field oscillations in the region $x>8$. The velocity distribution of the leptons in this region is, however, not a Maxwellian. Therefore the plasma contains free energy that can be released by a collisionless instability. The thermal anisotropy contained in the total lepton velocity distribution $f_t(v_x, v_y)$ and measured in the rest frame of the downstream plasma can be estimated as 
\begin{equation}
A =  \frac{\int f_t(v_x,v_y) {(v_x-p_0/2m_e)}^2 dv_x dv_y}{\int f_t(v_x,v_y) v_y^2 dv_x dv_y}-1.
\end{equation} 
A value $A=0$ would say that the thermal energy in the $x$ direction equals that in the $y$ direction, which would imply that there is no thermal anisotropy. We obtain the value $A\approx 6$ from the data shown in Fig. \ref{vxvy}(c). Such a large anisotropy value results in the Weibel instability in its original form \citep{Weibel59,Morse69}. 

\begin{figure}
{\includegraphics[width=0.5\columnwidth]{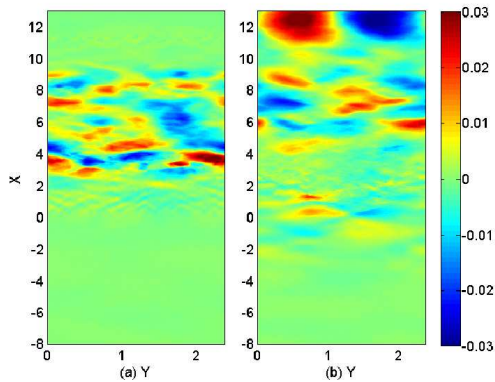}}
\caption{Panels (a) and (b) show the out-of-plane component $B_z$ of the magnetic field at the times $t_3 = 58.8$ and $t_4 = 120$, respectively. The color scale is the same for both panels. For comparison: The downstream region at $t_3 = 58.8$ is enclosed by shocks at $x\approx 3$ and $x\approx 8$, while the correctly resolved shock at $t_4 = 120$ is located at $x\approx 5$.}\label{Magnetic}
\end{figure}

Figure \ref{Magnetic} confirms that a magnetic field has grown in the downstream region. 
Strong magnetic fields with approximately the same peak amplitude are present at both times. The amplitude of the magnetic field exceeds that observed in Fig. \ref{FieldsStep1836}(c, f) by a factor 3 and it equals that of the electric field in the given normalization in Fig. \ref{FieldStep297}(a,b). Nevertheless, the magnetic force, which is acting on a lepton that is moving at the speed of $0.1$ relative to these field patches, will still be an order of magnitude weaker than the electric force in the shock transition layer. 

The strong downstream magnetic fields are not correlated with electric field structures (See Figs. \ref{FieldsStep144}(a, b) and \ref{FieldStep297}(a,b)) and they are thus not driven by an oblique mode instability. The Weibel instability drives magnetowaves with a negligible electric field and with the same magnetic field direction as the one observed here and this instability is thus compatible with the simulation data. The Weibel instability yields magnetic fields with an energy density that can reach up to 10\% of the thermal energy for strong temperature anisotropies \citep{Morse71,Stockem09}. The velocity spread in Fig. \ref{vxvy}(c) corresponds to a lepton temperature of the order of 1 keV or $10^7$ K. The magnetic pressure of a field $B_z = 0.03$ is thus a few percent of the cumulative thermal pressure of electrons and positrons and within the range that is accessible to the Weibel instability. 

We can compare the maximum size of the magnetic patches in Fig. \ref{Magnetic} to the gyroradius of an electron in that field. The magnetic field $\tilde{B}_z$ and the collision speed $\tilde{v}_0$ in physical units can be calculated from the normalized ones via $B_z = e \tilde{B}_z / m_e \omega_p$ and $\tilde{v}_0 = v_0 c$. The gyroradius of an electron that moves at the speed $v_0/2$ relative to the stationary perpendicular magnetic field, which is normalized to $\lambda_s = c/\omega_p$, is then $r_g / \lambda_s = v_0 / 2 B_z$. Taking $v_0 = 0.2$ and $B_z = 0.03$ gives $r_g \approx 3 \lambda_s$, which is about three times the coherence scale of the largest magnetic field patches. The magnetic field patches are also not stationary on the time scale needed to perform a gyro-orbit. This time equals for a maximum amplitude $B_{max} = 0.03$ in our normalization $2\pi / B_{max} \approx 200$, which exceeds the simulation time. The leptons can thus not complete a full gyro-orbit. The magnetic field will instead deflect leptons by a small angle that depends on where the lepton entered the patch and on how long it stayed inside the patch. The magnetic field will thus scatter the leptons of the directed beam in Fig. \ref{vxvy}(c). The repeated scattering of the leptons will eventually thermalize their distribution.  

\section{Summary}

We have examined the formation and the initial evolution of a non-relativistic leptonic shock. The shock was created by letting two spatially uniform clouds of equally dense electrons and positrons collide at a relative speed of 0.2c. The absence of binary collisions implied that both clouds initially interpenetrated and formed an overlap layer. The two-stream instability grew in this overlap layer. The nonlinear saturation of the growing waves heated up the plasma in the overlap layer and transformed it into a downstream region that was enclosed by two shocks. We followed the evolution of one of them.

Some of the hot downstream leptons escaped upstream and interacted with the inflowing upstream plasma. Nonlinear and predominantly electrostatic phase space vortices formed, which mediated the shock transition layer. These structures are unstable \citep{Wharton68,Berk67,Morse69} and their collapse scattered and heated the leptons and gave rise to electric field flucutations. The interaction of electric field fluctuations and charged particles has a similar effect as binary collisions between particles \citep{Dum78,Dieckmann06b,Bale02,Baalrud09,Bret15} with respect to the thermalization of the inflowing plasma and this interaction contributed to the particle heating by the shock.

A spatially confined region formed, in which the plasma density exceeded the cumulative density of the colliding clouds. We found the compression factor 3, which is expected for a strong shock. However, the electrostatic shock transition layer could not fully thermalize the upstream plasma that crossed it. The residual thermal anisotropy of the downstream plasma drove magnetowaves via the instability proposed by \citet{Weibel59} in its original form. The magnetic fields were weak and spatially inhomogeneous. Nevertheless, the leptons that would enter these field patches would experience a small-angle deflection by the magnetic field. 

The magnitude of the deflection angle depends on the time the particle needs to cross the magnetic patch. Repeated deflections will thus randomize the particle paths. This randomization will result in a thermalization of the downstream plasma on a time scale that exceeds by far the one accessible to our simulation. We note that the Weibel instability grows already at low values of the thermal anisotropy \citep{Morse69}. Magnetic field patches will thus grow and scatter particles until the lepton population has reached a thermal equilibrium. 

The motivation of our work has been to better understand the properties of the internal shocks of microquasar jets. The initial temperature, which we gave to the leptons, and the shock speed are, however, lower than the values we may find close to the internal shocks of microquasars jets. We chose these low values to make the waves and plasma structures in the shock transition layer quasi-electrostatic. We will study in future work larger non-relativistic collision speeds and consider the effects of a larger initial temperature and determine up to which values the shocks resemble the one we have examined here. 

We will also study in more detail the spectrum of the unstable waves that can grow in a pair plasma that consists of a cool beam immersed in a hot background population. Our simulation showed the presence of electrostatic waves in such a plasma. We proposed that such waves could be driven by an instability, which is similar to the electron acoustic instability \cite{Gary87}, and we have to determine if a similar instability exists in the pair plasma we considered here.

\textbf{Acknowledgements:} Grant ENE2013-45661-C2-1-P from the Ministerio de Educaci\'on y Ciencia, Spain and Grant PEII-2014-008-P from the Junta de Comunidades de Castilla-La Mancha. Computer time and support was provided by the HPC2N and by the Swedish National Infrastructure for Computing (SNIC) through the grant SNIC2015-1-305.

\bibliographystyle{jpp}
\bibliography{Manuscript}

\begin{thebibliography}{49}
\expandafter\ifx\csname natexlab\endcsname\relax\def\natexlab#1{#1}\fi
\def\au#1{#1} \def\ed#1{#1} \def\yr#1{#1}\def\at#1{#1}\def\jt#1{\textit{#1}}
  \def\bt#1{#1}\def\bvol#1{\textbf{#1}} \def\vol#1{#1} \def\pg#1{#1}
  \def\publ#1{#1}\def\arxiv#1{#1}\def\org#1{#1}\def\st#1{\textit{#1}}

\bibitem[Arber {\em et~al.\/}(2015)Arber, Bennett, Brady, Lawrence-Douglas,
  Ramsay, Sircombe, Gillies, Evans, Schmitz, Bell \& Ridgers]{Arber15}
{\sc \au{Arber, T.~D.}, \au{Bennett, K.}, \au{Brady, C.~S.},
  \au{Lawrence-Douglas, A.}, \au{Ramsay, M.~G.}, \au{Sircombe, N.~J.},
  \au{Gillies, P.}, \au{Evans, R.~G.}, \au{Schmitz, H.}, \au{Bell, A.~R.} \&
  \au{Ridgers, C.~P.}} \yr{2015}  \at{Contemporary particle-in-cell approach to
  laser-plasma modelling}.  \jt{Plasma Phys. Controll. Fusion}  \bvol{57},
  \pg{113001}.

\bibitem[Baalrud {\em et~al.\/}(2009)Baalrud, Callen \& Hegna]{Baalrud09}
{\sc \au{Baalrud, S.~D.}, \au{Callen, J.~D.} \& \au{Hegna, C.~C.}} \yr{2009}
  \at{Instability-enhanced collisional effects and langmuir's paradox}.
  \jt{Phys. Rev. Lett.}  \bvol{102},  \pg{245005}.

\bibitem[Bale {\em et~al.\/}(2002)Bale, Hull, Larson, Lin, Muschietti, Kellog,
  Goetz \& Monson]{Bale02}
{\sc \au{Bale, S.~D.}, \au{Hull, A.}, \au{Larson, D.~E.}, \au{Lin, R.~P.},
  \au{Muschietti, L.}, \au{Kellog, P.~J.}, \au{Goetz, K.} \& \au{Monson,
  S.~J.}} \yr{2002}  \at{Electrostatic turbulence and debye-scale structures
  associated with electron thermalization at collisionless shocks}.
  \jt{Astrophys. J.}  \bvol{575},  \pg{L25}.

\bibitem[Berk \& Roberts(1967)]{Berk67}
{\sc \au{Berk, H.~L.} \& \au{Roberts, K.~V.}} \yr{1967}  \at{Nonlinear study of
  vlasov's equation for a special class of distribution functions}.  \jt{Phys.
  Fluids}  \bvol{10},  \pg{1595}.

\bibitem[Brainerd(2000)]{Brainerd00}
{\sc \au{Brainerd, J.~J.}} \yr{2000}  \at{A plasma instability theory of
  gamma-ray burst emission}.  \jt{Astrophys. J.}  \bvol{538},  \pg{628--637}.

\bibitem[Bret(2015)]{Bret15}
{\sc \au{Bret, A.}} \yr{2015}  \at{Collisional behaviors of astrophysical
  collisionless plasmas}.  \jt{J. Plasma Phys.}  \bvol{81},  \pg{455810202}.

\bibitem[Bret \& Deutsch(2005)]{Bret2005}
{\sc \au{Bret, A.} \& \au{Deutsch, C.}} \yr{2005}  \at{Mixed two-stream
  filamentation modes in a collisional plasma}.  \jt{Phys. Plasmas}  \bvol{12},
   \pg{082704}.

\bibitem[Bret {\em et~al.\/}(2008)Bret, Gremillet, Benisti \& Lefebvre]{Bret08}
{\sc \au{Bret, A.}, \au{Gremillet, L.}, \au{Benisti, D.} \& \au{Lefebvre, E.}}
  \yr{2008}  \at{Exact relativistic kinetic theory of an electron-beam-plasma
  system: Hierarchy of the competing modes in the system-parameter space}.
  \jt{Phys. Rev. Lett.}  \bvol{100},  \pg{205008}.

\bibitem[Bret {\em et~al.\/}(2010)Bret, Gremillet \& Dieckmann]{Bret10}
{\sc \au{Bret, A.}, \au{Gremillet, L.} \& \au{Dieckmann, M.~E.}} \yr{2010}
  \at{Multidimensional electron beam-plasma instabilities in the relativistic
  regime}.  \jt{Phys. Plasmas}  \bvol{17},  \pg{120501}.

\bibitem[Bret {\em et~al.\/}(2013)Bret, Stockem, Fiuza, Ruyer, Gremillet,
  Narayan \& Silva]{Bret2013}
{\sc \au{Bret, A.}, \au{Stockem, A.}, \au{Fiuza, F.}, \au{Ruyer, C.},
  \au{Gremillet, L.}, \au{Narayan, R.} \& \au{Silva, L.~O.}} \yr{2013}
  \at{Collisionless shock formation, spontaneous electromagnetic fluctuations,
  and streaming instabilities}.  \jt{Phys. Plasmas}  \bvol{20},  \pg{042102}.

\bibitem[Bridle \& Perley(1984)]{Bridle84}
{\sc \au{Bridle, A.~H.} \& \au{Perley, R.~A.}} \yr{1984}  \at{Extragalactic
  radio jets}.  \jt{Ann. Rev. Astron. Astrophys.}  \bvol{22},  \pg{319}.

\bibitem[Califano {\em et~al.\/}(1998)Califano, Prandi, Pegoraro \&
  Bulanov]{Califano}
{\sc \au{Califano, F.}, \au{Prandi, R.}, \au{Pegoraro, F.} \& \au{Bulanov,
  S.~V.}} \yr{1998}  \at{Nonlinear filamentation instability driven by an
  inhomogeneous current in a collisionless plasma}.  \jt{Phys. Rev. E}
  \bvol{58},  \pg{7837}.

\bibitem[Chang {\em et~al.\/}(2008)Chang, Spitkovsky \& Arons]{Chang08}
{\sc \au{Chang, P.}, \au{Spitkovsky, A.} \& \au{Arons, J.}} \yr{2008}
  \at{Long-term evolution of magnetic turbulence in relativistic collisionless
  shocks: Electron-positron plasmas}.  \jt{Astrophys. J.}  \bvol{674},
  \pg{378}.

\bibitem[Dieckmann {\em et~al.\/}(2006{\natexlab{{\em a\/}}})Dieckmann, Drury
  \& Shukla]{Dieckmann06b}
{\sc \au{Dieckmann, M.~E.}, \au{Drury, L.~O.} \& \au{Shukla, P.~K.}}
  \yr{2006{\natexlab{{\em a\/}}}}  \at{On the ultrarelativistic two-stream
  instability, electrostatic turbulence and brownian motion}.  \jt{New J.
  Phys.}  \bvol{8},  \pg{40}.

\bibitem[Dieckmann {\em et~al.\/}(2006{\natexlab{{\em b\/}}})Dieckmann,
  Frederiksen, Bret \& Shukla]{Dieckmann06}
{\sc \au{Dieckmann, M.~E.}, \au{Frederiksen, J.~T.}, \au{Bret, A.} \&
  \au{Shukla, P.~K.}} \yr{2006{\natexlab{{\em b\/}}}}  \at{Evolution of the
  fastest-growing relativistic mixed mode instability driven by a tenuous
  plasma beam in one and two dimensions}.  \jt{Phys. Plasmas}  \bvol{13},
  \pg{112110}.

\bibitem[Dum(1978)]{Dum78}
{\sc \au{Dum, C.~T.}} \yr{1978}  \at{Anomalous heating by ion sound
  turbulence}.  \jt{Phys. Fluids}  \bvol{21},  \pg{945}.

\bibitem[Fabian \& Rees(1979)]{Fabian79}
{\sc \au{Fabian, A.~C.} \& \au{Rees, M.~J.}} \yr{1979}  \at{Ss433-double jet in
  action}.  \jt{Mon. Not. R. Astron. Soc.}  \bvol{187},  \pg{P13}.

\bibitem[Gary(1987)]{Gary87}
{\sc \au{Gary, S.~P.}} \yr{1987}  \at{The electron electron acoustic
  instability}.  \jt{Phys. Fluids}  \bvol{30},  \pg{2745}.

\bibitem[Haruki \& Sakai(2003)]{Haruki03}
{\sc \au{Haruki, T} \& \au{Sakai, J.~I.}} \yr{2003}  \at{Generation of magnetic
  field and electrostatic shock wave driven by counterstreaming pair plasmas}.
  \jt{Phys. Plasmas}  \bvol{10},  \pg{392--397}.

\bibitem[Jaroschek {\em et~al.\/}(2004)Jaroschek, Lesch \&
  Treumann]{Jaroschek04}
{\sc \au{Jaroschek, C.~H.}, \au{Lesch, H.} \& \au{Treumann, R.~A.}} \yr{2004}
  \at{Self-consistent diffusive lifetimes of weibel magnetic fields in
  gamma-ray bursts}.  \jt{Astrophys. J.}  \bvol{616},  \pg{1065--1071}.

\bibitem[Kaiser {\em et~al.\/}(2000)Kaiser, Sunyaev \& Spruit]{Kaiser00}
{\sc \au{Kaiser, C.~R.}, \au{Sunyaev, R.} \& \au{Spruit, H.~C.}} \yr{2000}
  \at{Internal shock model for microquasars}.  \jt{Astron. Astrophys.}
  \bvol{356},  \pg{975}.

\bibitem[Kazimura {\em et~al.\/}(1998)Kazimura, Sakai, Neubert \&
  Bulanov]{Kazimura98}
{\sc \au{Kazimura, Y.}, \au{Sakai, J.~I.}, \au{Neubert, T.} \& \au{Bulanov,
  S.~V.}} \yr{1998}  \at{Generation of a small-scale quasi-static magnetic
  field and fast particles during the collision of electron-positron plasma
  clouds}.  \jt{Astrophys. J.}  \bvol{498},  \pg{L183--L186}.

\bibitem[Kono {\em et~al.\/}(1980)Kono, Skoric \& ter Haar]{Kono80}
{\sc \au{Kono, M.}, \au{Skoric, M.~M.} \& \au{ter Haar, D.}} \yr{1980}
  \at{Ponderomotive force in a dispersive medium in a variable
  electromagnetic-field}.  \jt{Phys. Rev. Lett.}  \bvol{45},  \pg{1629}.

\bibitem[Kulkarni {\em et~al.\/}(1998)Kulkarni, Frail, Wieringa, Ekers, Sadler,
  Wark, Higdon, Phinney \& Bloom]{Kulkarni98}
{\sc \au{Kulkarni, S.~R.}, \au{Frail, D.~A.}, \au{Wieringa, M.~H.}, \au{Ekers,
  R.~D.}, \au{Sadler, E.~M.}, \au{Wark, R.~M.}, \au{Higdon, J.~L.},
  \au{Phinney, E.~S.} \& \au{Bloom, J.~S.}} \yr{1998}  \at{Radio emission from
  the unusual supernova 1998bw and its association with the gamma-ray burst of
  25 april 1998}.  \jt{Nature}  \bvol{395},  \pg{663}.

\bibitem[Landau(1946)]{Landau46}
{\sc \au{Landau, L.~D.}} \yr{1946}  \at{On the vibrations of the electronic
  plasma}.  \jt{J. Phys. (Moscow)}  \bvol{10},  \pg{25}.

\bibitem[Marcowith {\em et~al.\/}(2016)Marcowith, Bret, Bykov, Dieckmann,
  Drury, Lembege, Lemoine, Morlino, Murphy, Pelletier, Plotnikov, Reville,
  Riquelme, Sironi \& Novo]{Marcowith16}
{\sc \au{Marcowith, A.}, \au{Bret, A.}, \au{Bykov, A.}, \au{Dieckmann, M.~E.},
  \au{Drury, L.~O.}, \au{Lembege, B.}, \au{Lemoine, M.}, \au{Morlino, G.},
  \au{Murphy, G.}, \au{Pelletier, G.}, \au{Plotnikov, I.}, \au{Reville, B.},
  \au{Riquelme, M.}, \au{Sironi, L.} \& \au{Novo, A.~S.}} \yr{2016}  \at{The
  microphysics of collisionless shock waves}.  \jt{Rep. Prog. Phys.}
  \bvol{79},  \pg{046901}.

\bibitem[Margon(1984)]{Margon84}
{\sc \au{Margon, B.}} \yr{1984}  \at{Observations of ss-433}.  \jt{Ann. Rev.
  Astron. Astrophys.}  \bvol{22},  \pg{507}.

\bibitem[Medvedev {\em et~al.\/}(2005)Medvedev, Fiore, Fonseca, Silva \&
  Mori]{Medvedev05}
{\sc \au{Medvedev, M.~V.}, \au{Fiore, M.}, \au{Fonseca, R.~A.}, \au{Silva,
  L.~O.} \& \au{Mori, W.~B.}} \yr{2005}  \at{Long-time evolution of magnetic
  fields in relativistic gamma-ray burst shocks}.  \jt{Astrophys. J.}
  \bvol{618},  \pg{L75}.

\bibitem[Medvedev \& Loeb(1999)]{Medvedev99}
{\sc \au{Medvedev, M.~V.} \& \au{Loeb, A.}} \yr{1999}  \at{Generation of
  magnetic fields in the relativistic shock of gamma-ray burst sources}.
  \jt{Astrophys. J.}  \bvol{526},  \pg{697--706}.

\bibitem[Miller-Jones {\em et~al.\/}(2005)Miller-Jones, McCormick, P., Spencer,
  Muxlow \& Pooley]{Miller05}
{\sc \au{Miller-Jones, J. C.~A.}, \au{McCormick, D.~G.}, \au{P., Fender.~R.},
  \au{Spencer, R.~E.}, \au{Muxlow, T. W.~B.} \& \au{Pooley, G.~G.}} \yr{2005}
  \at{Multiple relativistic outbursts of grs 1915+105: radio emission and
  internal shocks}.  \jt{Mon. Not. R. Astron. Soc.}  \bvol{363},  \pg{867}.

\bibitem[Milosavljevic \& Nakar(2006)]{Milos06}
{\sc \au{Milosavljevic, M.} \& \au{Nakar, E.}} \yr{2006}  \at{Weibel filament
  decay and thermalization in collisionless shocks and gamma-ray burst
  afterglows}.  \jt{Astrophys. J.}  \bvol{641},  \pg{978--983}.

\bibitem[Morse \& Nielson(1969)]{Morse69}
{\sc \au{Morse, R.~L.} \& \au{Nielson, C.~W.}} \yr{1969}  \at{One-, two- and
  three-dimensional numerical simulation of 2-beam plasmas}.  \jt{Phys. Rev.
  Lett.}  \bvol{23},  \pg{1087}.

\bibitem[Morse \& Nielson(1971)]{Morse71}
{\sc \au{Morse, R.~L.} \& \au{Nielson, C.~W.}} \yr{1971}  \at{Numerical
  simulation of the weibel instability in one and two dimensions}.  \jt{Phys.
  Fluids}  \bvol{14},  \pg{830}.

\bibitem[O'Neil(1965)]{ONeil65}
{\sc \au{O'Neil, T.~M.}} \yr{1965}  \at{Collisionless damping of nonlinear
  plasma oscillations}.  \jt{Phys. Fluids}  \bvol{8},  \pg{2255}.

\bibitem[Oppenheim {\em et~al.\/}(1999)Oppenheim, Newman \&
  Goldman]{Oppenheim99}
{\sc \au{Oppenheim, M.}, \au{Newman, D.~L.} \& \au{Goldman, M.~V.}} \yr{1999}
  \at{Evolution of electron phase-space holes in a 2d magnetized plasma}.
  \jt{Phys. Rev. Lett.}  \bvol{83},  \pg{2344}.

\bibitem[Rees(1978)]{Rees78}
{\sc \au{Rees, M.~J.}} \yr{1978}  \at{M87 jet - internal shocks in a plasma
  beam}.  \jt{Mon. Not. R. Astron. Soc.}  \bvol{184},  \pg{61}.

\bibitem[Sakai {\em et~al.\/}(2000)Sakai, Nakayama, Kazimura \&
  Bulanov]{Sakai00}
{\sc \au{Sakai, J.}, \au{Nakayama, T.}, \au{Kazimura, Y.} \& \au{Bulanov, S.}}
  \yr{2000}  \at{Magnetic field generation and subsequent field dissipation
  with plasma heating in relativistic streaming pair plasmas}.  \jt{J. Phys.
  Soc. Jpn.}  \bvol{69},  \pg{2503--2513}.

\bibitem[Silva {\em et~al.\/}(2003)Silva, Fonseca, Tonge, Dawson, Mori \&
  Medvedev]{Silva03}
{\sc \au{Silva, L.~O.}, \au{Fonseca, R.~A.}, \au{Tonge, J.~W.}, \au{Dawson,
  J.~M.}, \au{Mori, W.~B.} \& \au{Medvedev, M.~V.}} \yr{2003}
  \at{Interpenetrating plasma shells: Near-equipartition magnetic field
  generation and nonthermal particle acceleration}.  \jt{Astrophys. J.}
  \bvol{596},  \pg{L121--L124}.

\bibitem[Silva {\em et~al.\/}(2002)Silva, Fonseca, Tonge, Mori \&
  Dawson]{Silva2002}
{\sc \au{Silva, L.~O.}, \au{Fonseca, R.~A.}, \au{Tonge, J.~W.}, \au{Mori,
  W.~B.} \& \au{Dawson, J.~M.}} \yr{2002}  \at{On the role of the purely
  transverse weibel instability in fast ignitor scenarios}.  \jt{Phys. Plasmas}
   \bvol{9},  \pg{2458}.

\bibitem[Sironi \& Giannios(2014)]{Sironi14}
{\sc \au{Sironi, L.} \& \au{Giannios, D.}} \yr{2014}  \at{Relativistic pair
  beams from tev blazars: A source of reprocessed gev emission rather than
  intergalactic heating}.  \jt{Astrophys. J.}  \bvol{787},  \pg{49}.

\bibitem[Stockem {\em et~al.\/}(2008)Stockem, Dieckmann \&
  Schlickeiser]{Stockem08}
{\sc \au{Stockem, A.}, \au{Dieckmann, M.~E.} \& \au{Schlickeiser, R.}}
  \yr{2008}  \at{Suppression of the filamentation instability by a flow-aligned
  magnetic field: testing the analytic threshold with pic simulations}.
  \jt{Plasma Phys. Controll. Fusion}  \bvol{50},  \pg{025002}.

\bibitem[Stockem {\em et~al.\/}(2009)Stockem, Dieckmann \&
  Schlickeiser]{Stockem09}
{\sc \au{Stockem, A.}, \au{Dieckmann, M.~E.} \& \au{Schlickeiser, R.}}
  \yr{2009}  \at{Pic simulations of the thermal anisotropy-driven weibel
  instability: field growth and phase space evolution upon saturation}.
  \jt{Plasma Phys. Controll. Fusion}  \bvol{51},  \pg{075014}.

\bibitem[Stockem {\em et~al.\/}(2014)Stockem, Grismayer, Fonseca \&
  Silva]{Stockem14}
{\sc \au{Stockem, A.}, \au{Grismayer, T.}, \au{Fonseca, R.~A.} \& \au{Silva,
  L.~O.}} \yr{2014}  \at{Electromagnetic field generation in the downstream of
  electrostatic shocks due to electron trapping}.  \jt{Phys. Rev. Lett.}
  \bvol{113},  \pg{105002}.

\bibitem[Trigo {\em et~al.\/}(2013)Trigo, Miller-Jones, Migliari, Broderick \&
  Tzioumis]{Trigo13}
{\sc \au{Trigo, M.~D.}, \au{Miller-Jones, J.~C.~A.}, \au{Migliari, S.},
  \au{Broderick, J.~W.} \& \au{Tzioumis, T.}} \yr{2013}  \at{Baryons in the
  relativistic jets of the stellar-mass black-hole candidate 4u 1630-47}.
  \jt{Nature}  \bvol{504},  \pg{260--262}.

\bibitem[Umeda {\em et~al.\/}(2003)Umeda, Omura, Yoon, Gaelzer \&
  Matsumoto]{Umeda03}
{\sc \au{Umeda, T.}, \au{Omura, Y.}, \au{Yoon, P.~H.}, \au{Gaelzer, R.} \&
  \au{Matsumoto, H.}} \yr{2003}  \at{Harmonic langmuir waves. iii. vlasov
  simulation}.  \jt{Phys. Plasmas}  \bvol{10},  \pg{382}.

\bibitem[Weibel(1959)]{Weibel59}
{\sc \au{Weibel, E.~S.}} \yr{1959}  \at{Spontaneously growing transverse waves
  in a plasma due to an anisotropic velocity distribution}.  \jt{Phys. Rev.
  Lett.}  \bvol{2},  \pg{83}.

\bibitem[Wharton {\em et~al.\/}(1968)Wharton, Malmberg \& O'Neil]{Wharton68}
{\sc \au{Wharton, C.~B.}, \au{Malmberg, J.~H.} \& \au{O'Neil, T.~M.}} \yr{1968}
   \at{Nonlinear effects of large-amplitude plasma waves}.  \jt{Phys. Fluids}
  \bvol{11},  \pg{1761}.

\bibitem[Woosley \& Bloom(2006)]{Woosley06}
{\sc \au{Woosley, S.~E.} \& \au{Bloom, J.~S.}} \yr{2006}  \at{The
  supernova-gamma-ray burst connection}.  \jt{Ann. Rev. Astron. Astrophys.}
  \bvol{{44}},  \pg{507--556}.

\bibitem[Zel'Dovich {\em et~al.\/}(1967)Zel'Dovich, Raizer, Probstein \&
  Hayes]{Zeldovich67}
{\sc \au{Zel'Dovich, Ya.~B.}, \au{Raizer, Yuri~P.}, \au{Probstein, Ronald~F.}
  \& \au{Hayes, Wallace~D.}} \yr{1967} {\em Physics of shock waves and
  high-temperature hydrodynamic phenomena\/}.  \publ{London, New York, Sydney:
  Academic press}.

\end{thebibliography}


\begin{thebibliography}{14}
\expandafter\ifx\csname natexlab\endcsname\relax\def\natexlab#1{#1}\fi

\bibitem[Batchelor(1971)]{Batchelor59}
{\sc Batchelor, G.~K.} 1971 Small-scale variation of convected quantities like
  temperature in turbulent fluid. part 1. general discussion and the case of
  small conductivity. {\em J.~Fluid Mech.\/} {\bf 5}, 113--133.

\bibitem[Brownell \& Su(2004)]{Brownell04}
{\sc Brownell, C.~J. \& Su, L.~K.} 2004 Planar measurements of differential
  diffusion in turbulent jets. {\em AIAA Paper 2004-2335\/}.

\bibitem[Brownell \& Su(2007)]{Brownell07}
{\sc Brownell, C.~J. \& Su, L.~K.} 2007 Scale relations and spatial spectra in
  a differentially diffusing jet. {\em AIAA Paper 2007-1314\/}.

\bibitem[Dennis(1985)]{Dennis85}
{\sc Dennis, S. C.~R.} 1985 {Compact explicit finite difference approximations
  to the Navier--Stokes equation}. In {\em Ninth Intl Conf. on Numerical
  Methods in Fluid Dynamics\/} (ed. Soubbaramayer \& J.~P. Boujot), {\em
  Lecture Notes in Physics\/}, vol. 218, pp. 23--51. Springer.

\bibitem[Hwang \& Tuck(1970)]{Hwang70}
{\sc Hwang, L.-S. \& Tuck, E.~O.} 1970 On the oscillations of harbours of
  arbitrary shape. {\em J.~Fluid Mech.\/} {\bf 42}, 447--464.

\bibitem[Koch(1983)]{Koch83}
{\sc Koch, W.} 1983 Resonant acoustic frequencies of flat plate cascades. {\em
  J.~Sound Vib.\/} {\bf 88}, 233--242.

\bibitem[Lee(1971)]{Lee71}
{\sc Lee, J.-J.} 1971 Wave-induced oscillations in harbours of arbitrary
  geometry. {\em J.~Fluid Mech.\/} {\bf 45}, 375--394.

\bibitem[Linton \& Evans(1992)]{Linton92}
{\sc Linton, C.~M. \& Evans, D.~V.} 1992 The radiation and scattering of
  surface waves by a vertical circular cylinder in a channel. {\em Phil.\
  Trans.\ R. Soc.\ Lond.\/} {\bf 338}, 325--357.

\bibitem[Martin(1980)]{Martin80}
{\sc Martin, P.~A.} 1980 On the null-field equations for the exterior problems
  of acoustics. {\em Q.~J. Mech.\ Appl.\ Maths\/} {\bf 33}, 385--396.

\bibitem[Miller(1991)]{Miller91}
{\sc Miller, P.~L.} 1991 Mixing in high schmidt number turbulent jets. PhD
  thesis, California Institute of Technology.

\bibitem[Rogallo(1981)]{Rogallo81}
{\sc Rogallo, R.~S.} 1981 Numerical experiments in homogeneous turbulence. {\em
  Tech. Rep.\/} 81835. NASA Tech.\ Mem.

\bibitem[Ursell(1950)]{Ursell50}
{\sc Ursell, F.} 1950 Surface waves on deep water in the presence of a
  submerged cylinder i. {\em Proc.\ Camb.\ Phil.\ Soc.\/} {\bf 46}, 141--152.

\bibitem[{van Wijngaarden}(1968)]{Wijngaarden68}
{\sc {van Wijngaarden}, L.} 1968 On the oscillations near and at resonance in
  open pipes. {\em J.~Engng Maths\/} {\bf 2}, 225--240.

\bibitem[Worster(1992)]{Worster92}
{\sc Worster, M.~G.} 1992 {The dynamics of mushy layers}. In {\em In
  Interactive dynamics of convection and solidification\/} (ed. S.~H. Davis,
  H.~E. Huppert, W.~Muller \& M.~G. Worster), pp. 113--138. Kluwer.

\end{thebibliography}

\end{document}